\documentclass{elsart}
\usepackage{amssymb}
\usepackage{graphics}
\begin{document}

\begin{frontmatter}



\title{Long-time fidelity and chaos for a kicked nonlinear oscillator system}


\author{A. Kowalewska-Kud{\l}aszyk\thanksref{ak}}
\thanks[ak]{annakow@amu.edu.pl}
\author{J.K. Kalaga}
\author{W. Leo\'nski\thanksref{wl}}
\thanks[wl]{wleonski@amu.edu.pl}
\address{Nonlinear Optics Division, Institute of Physics, Adam Mickiewicz
 University, Umultowska 85, 61-614 Pozna\'n, Poland}

\begin{abstract}
We deal with a system comprising a nonlinear (Kerr-like) oscillator excited by a series of ultra-short external pulses. We introduce the fidelity-based entropic parameter that can be used as an indicator of quantum chaos. Moreover, we propose to use the fidelity-like parameter comprising the information about the mean number of photons in the system. We shall concentrate on the long-time behaviour of the parameters discussed, showing that for deep chaos cases the  quantum fidelities behave chaotically in the classical sense despite their strictly quantum character. 
\end{abstract}

\end{frontmatter}

\section{Introduction}
There is still a great interest in quantum physics systems that can
demonstrate chaotic motion. It is of special interest to find
strict quantum indicators that would determine the frontiers between
regular and chaotic regions in the system's dynamics.
In classical dynamics one can talk about a chaotic motion when the system
(whose dynamics is described by the nonlinear equations) 
is sensitive to initial conditions. The orbits in the phase space are
unstable and errors in the initial data grow exponentially  and
consequently, the final  state of the system is unpredictable. Methods
for  analysing such a situation are well developed. For instance, one can numerically determine the
chaotic behaviour of the system analysing the Lyapunov exponents.
For quantum systems, due to the linearity of the Schr{\"{o}}dinger
equation and consequently, the fact that the dynamics is governed by
the unitary evolution operator, small changes in the initial state do
not cause indefinite changes in the final state of the quantum
system. Therefore, standard methods used for problems of
classical  chaos cannot be used for such cases. In consequence, there is a need for finding quantum
signatures that would be associated with chaos appearing in quantum
systems and hence, 
it is of special interest to analyse the quantum counterparts of
the physical systems which in their classical version have both regular
and chaotic domains (for instance see \cite{HKS87,DES92}). 
 
At this point one should mention various 
approaches to the quantum chaos signatures that have been
proposed \cite{H92,P95}. The most often used is that based upon the
correspondence  between the statistics of eigenvalues and eigenvectors
of  quantum states (which in the classical limit behave chaotically) and
the  canonical ensembles of the random matrix theory RMT \cite{BGS84,BT77,I87,KMH88,HZ90}. 
Moreover, in the quantum information theory, while considering multiquibit systems,
the degree of entanglement between states can be used as a signature of  chaotic
behaviour of the system. It has been shown that bipartite entanglement in chaotic regions is enhanced,
whereas the pairwise entanglement in those regions is suppressed
\cite{WGSH04}. Additionally, the time averaged entangling power has also been
defined and  proposed as an indicator of the limit of chaos in the
quantum  system's dynamics \cite{WGSH04}.

On the other hand, considering quantum dynamics of systems behaving
chaotically  in the classical limit, the time varying fidelity between
quantum  states can be used as a signature of chaos
\cite{P84,WLT02,EWLC02}. The fidelity is a measure of stability of
time evolution of quantum states. It is also a standard quantity
allowing a measure of decoherence in quantum computations.
It has been proved that whenever a quantum system
begins to behave chaotically, the fidelity decreases exponentially
\cite{JP01, JSB01, PZ02, GPSZ06, GPS04}.
The rate of the fidelity decay has been a subject of much investigation and depending on the perturbation character and its strength various types of that decay have been identified. A thorough analysis of this problem can be found in \cite{GPSZ06}.
Generally, this character can be explained on the basis of random matrix theory \cite{GPS04}. Moreover, one should mention that 
some attempts to determine the fractal dimension of the fidelity between quantum states as an indicator of quantum chaos have also been made \cite{PM07}. 

It has been shown that the perturbation independent exponential decay governed by the Lyapunov exponent (of classical chaotic dynamics) is characteristic of the perturbation strength which is beyond the applicability of the perturbation theory.
In the perturbative regime (small perturbations) the fidelity decay is of Gaussian type. This type of fidelity decay is also characteristic of integrable and quasi-integrable classical dynamics.
Increasing the value of the perturbation strength one enters a golden rule regime of fidelity decay with the perturbation strength dependent slope of the exponential decay \cite{JSB01}. In general, the decay of quantum fidelity in quantum chaotic regions is slower than for integrable ones.

The main aim of the considerations presented in this paper is to analyse the 
long-time limit of the fidelity decay and the 
applicability of the strictly quantum parameter --
the fidelity based entropy -- for distinction between
regular and chaotic dynamics of the quantum system. 
We shall show that the long-time analysis of the fidelity-based parameters that are of strictly quantum nature, will be a good counterpart of the classical chaos indicators. 
As a model for our considerations we use the quantum analogue of a kicked Kerr
nonlinear oscillator (some features of which have been discussed in \cite{KKL04}), which in its classical dynamics exhibits both regular and chaotic motion \cite{L96}. 

\section{The model}
We consider a Kerr-like nonlinear oscillator (initially in the vacuum
state)  that is externally driven by a series of ultra short coherent pulses.
The system considered is also known as a kicked nonlinear oscillator
and its various aspects have been extensively discussed in numerous papers
(see for example \cite{L96, LT94, MH86, LDT97, GLSz98, SzGB93} {\it end the references quoted therein}).
It is also known that such an oscillator can demonstrate regular
dynamics as well as classically chaotic one. These two
types of the system's dynamics can be observed depending on the parameters used.

The system under consideration is described 
by the following Hamiltonian in the interaction picture:
\begin{equation}
\hat{H}=\hat{H}_{NL}+\hat{H}_{K}\,\,,
\end{equation}
where $\hat{H}_{NL}$ governs the evolution of the system between the two subsequent
pulses  and $\hat{H}_K$ -- during the infinitesimally short pulse:
\begin{eqnarray}
\hat{H}_{NL}&=&\frac{\chi}{2}\left(\hat{a}^+\right)^2\hat{a}^2\,\,,\label{eq1}\\
\hat{H}_{K}&=&\epsilon\left(\hat{a}^{+}+\hat{a}\right)\sum\limits_{k=1}^\infty\delta(t-kT)\,\, ,
\label{eq2}
\end{eqnarray}
where we use units $\hbar=1$. Operators $\hat{a}^+$ and $\hat{a}$ appearing here are the usual photon creation
and  annihilation operators respectively, $\chi$ describes the nonlinearity of the
oscillator (for a Kerr medium it is the third order susceptibility),
$\epsilon$  is the strength of the external pulses -- nonlinear system interaction, and $T$ is the time
between two subsequent pulses. Under the assumption that time $T$ exceeds
significantly the duration of a single pulse, we can model the series of
ultra-short coherent pulses by a series of Dirac-delta functions.
Our considerations are restricted to the case of the system
without damping processes, so, we can solve the problem using the
wave-function approach. It is also possible to include the damping to
the oscillator's dynamics but then the problem would need the density matrix
approach  and will be addressed elsewhere.

To obtain the evolution of the wave-function $|\Psi\rangle$ we need to
know the explicit form of the evolution operator. Therefore, we divide the
system's  evolution into two parts. First, we define the nonlinear
evolution  operator $\hat{U}_{NL}$ that would describe the dynamics between
two subsequent pulses and next, the operator $\hat{U}_K$ that would be
responsible  for the changes in the system during the action of the
ultra-short coherent pulse.
If T is the time between the two subsequent pulses and $\hat{H}_{NL}$
in  (\ref{eq1}) is the Hamiltonian describing the
system without the action of the external field, the
unitary  evolution operator corresponding to this period of time has the following form:
\begin{equation}
\hat{U}_{NL}=e^{-i\chi T\hat{n}(\hat{n}-1)}\,\,,
\label{eq3}
\end{equation}
where the operator $\hat{n}$ is the photon number operator. 
For the description of the evolution under the action of the
ultra-short external pulse we introduce the "kick" operator that can be expressed as:
\begin{equation}
\hat{U}_K=e^{-i\epsilon(\hat{a}^{+}+\hat{a})}\,\,.
\label{eq4}
\end{equation}

The evolution of the wave-function from the moment just after the $n$-th pulse
to  the moment just after the $(n+1)$-th one is then described by the action
of  these two operators, and after $k$-pulses the wave-function takes the form:
\begin{equation}
|\Psi_u(k)\rangle=\left(\hat{U}_{NL}\hat{U}_K\right)^k|\Psi(t=0)\rangle\,\, .
\label{eq5}
\end{equation}
In this way we are able to construct the map corresponding to the system's
quantum evolution and the mapping procedure is performed for the moments just
before the external pulses. Such calculations can be easily performed
numerically. 

\section{The fidelity}
It has been already shown that the character of the decay of the fidelity between quantum
states can be used as an indicator of the quantum chaos. It was used in this meaning by Peres \cite{P84}
and its applicability was developed by Weinstein \cite{WLT02} and
Emerson {\it et al.} \cite{EWLC02}.
They have shown that when the system's dynamics approaches the
chaotic region, the fidelity between especially chosen quantum states decays
in a characteristic way. They have discussed the states $|\Psi_u(n)\rangle$ 
evaluated in the unitary mapping procedure
and the ones $|\Psi_p(n)\rangle$ evaluated in the analogous
mapping but with some tiny perturbations, showing that the time evolution of fidelity can be used as an
indicator whether the quantum chaotic region has been achieved or not. \\
Thus, the fidelity between quantum states under consideration can be written as:
\begin{equation}
F(n)=|\langle\Psi_u(n)|\Psi_p(n)\rangle|\,\,,
\label{eq6}
\end{equation}
where the procedure for achieving the evolution of the unperturbed system
has already been given (see eq.(\ref{eq5})). Now, we have to describe
the unitary evolution of the perturbed system. The evolution operator for the system perturbed
corresponding to the interactions with the
external field is given by:
\begin{equation}
\hat{U}_{Kp}=e^{-i\left(\epsilon+
\Delta_{\epsilon}\right)\left(\hat{a}^{+}+\hat{a}\right)}\,\,,
\label{eq7}
\end{equation} 
where the parameter $\Delta_{\epsilon}\ll\epsilon$ describes the perturbation strength. Thus, we can write the fidelity in the following form:
\begin{equation}
F(t)=\left|\langle\Psi(t=0)|\left(\hat{U}^{+}_{Kp}\hat{U}^{+}_{NL}\right)^k
\left(\hat{U}_{NL}\hat{U}_K\right)^k|\Psi(t=0)\rangle\right|\,\,.
\label{eq8}
\end{equation}
Having expression (\ref{eq8}) it is possible to perform the
numerical calculations of the fidelity. 

First, we have to analyse the
dynamics of the system considered for various values of the strength
of the external field $\epsilon$. To determine whether the system (for a given value of the interaction strength $\epsilon$) is classically chaotic or not, we have plotted the
bifurcation diagrams for the mean energy of the classical counterpart of the system 
considered here. The bifurcation diagrams are plotted according to the procedure 
explained for instance in \cite{L96}.
To get it, first we write the solutions for the equations of motion for the annihilation and creation operators
evolving between the two subsequent pulses under the action of the unitary evolution operator
(\ref{eq3}). The solutions can be achieved in a simple analytical form, because in our model the damping is absent and consequently, the number of photons is preserved. This leads
to the formula:
\begin{equation}
\hat{a}(T)=e^{-i\chi\hat{n}T}\hat{a}\,\,\,.
\label{eq3a}
\end{equation}
The influence of the kicks (described 
by the action of the operator $\hat{U}_K$ -- eq.(\ref{eq4})) is reduced to the action of the shift operator onto (\ref{eq3a}). Hence, the annihilation operator just after $(k+1)$-th pulse can be expressed by the following recurrence relation:
\begin{equation}
\hat{a}_{k+1}=e^{-i\chi T(\hat{a}^{+}_{k}+i\epsilon)(\hat{a}_k-i\epsilon)}(\hat{a}_k-i\epsilon).
\label{eq3b}
\end{equation}
The final point of this procedure is to replace all the operators $\hat{a}^{+}(\hat{a})$ by the complex
numbers $\alpha^{\star}(\alpha)$, and  find the formula for the classical mean energy $|\alpha|^2$. 
Analysis of the bifurcation diagram for the mean classical energy shows that
for the nonlinearity parameter $\chi=1$ and the time between
subsequent pulses $T=\pi$ we have two regular regions 
(for $0<\epsilon<0.344$ and $0.356<\epsilon<0.47$).
For $\epsilon\approx 0.35$ we can observe a tiny chaotic region surrounded by regular ones, and for $\epsilon>0.47$ the system starts to behave chaotically again.

As our aim is to study the quantum counterpart of the classically chaotic system, 
our next objective is to study the strictly quantum parameter (the fidelity) in the quantum regions that are counterparts of those corresponding to the classically chaotic dynamics.
Therefore, we can analyse the fidelity (\ref{eq8}) for the values of
the external field strength that would correspond to the regular and chaotic
behaviours of the classical counterpart of the quantum system discussed. 
When the quantum system is in the region corresponding to the classical regular
dynamics, the fidelity changes periodically with time and these
oscillations are not perturbed (Fig.1). These oscillations are of periodic character determined by a single frequency. 
Recurrences of the fidelity (rebuilding the initial state $F=1$) depend on the perturbation strength solely and for the parameters used for creating Fig.1, the fidelity oscillations are of the period $T_{rec}=3333$ $[kicks]$ -- see Fig.2a.
Additionally, we can see that $T_{rec}$ decreases with increasing value of perturbation $\Delta_\epsilon$. For instance, for $\Delta_\epsilon=0.005$ the fidelity recurres every 630 external kicks (Fig.2b) whereas for $\Delta_\epsilon=0.08$ about 40 kicks are necessary for the fidelity recurrence -- Fig.2c.
In further considerations we will use the perturbation strength $\Delta_\epsilon=0.001$ and, for this value the orbits in phase space of the classical counterpart of our system are not significantly influenced. Therefore, for this case we can expect a Gaussian fidelity decay \cite{WH05}. It is clearly seen even for $\Delta_\epsilon=0.005$  (Fig.3) where we have plotted a logarithm of the fidelity versus time square - a linear approximation therefore is fully justified. From \cite{WH05} it is known that the rate of this decay is proportional to $\Delta_\epsilon^2$.

The character of fidelity oscillations and decay does not change when the field strength is sufficient to achieve the first chaotic region of dynamics in a classical system ($\epsilon\approx 0.35$).
The first changes in the fidelity evolution character are visible when we are closer to the main chaotic region. For instance, for $\epsilon\approx 0.385$ a slow modulation of the oscillations of previously determined frequency can be seen (Fig.4a). 
Therefore, we can see that when $\epsilon$ increases to such a value that the system approaches the region of chaotic dynamics, other frequencies start to play an important role.

Finally, for the perturbation strength which causes the chaotic dynamics of the classical system we can observe distinguishable changes in quantum fidelity behaviour --- Fig.4b. Apart from the initial decay (whose character in fact is not a subject of this paper) we can see well pronounced irregular changes. 
In fact, for this case, the fidelity has several dominant frequency components and some other that are less pronounced.
In further parts of this paper we will concentrate on these irregularities as possible indicators of chaotic dynamics of the quantum system analysed. 

As many papers dealing with quantum chaos have been devoted to the analysis of the character of  fidelity decay we will shortly focus on this problem.
When analysing the character of initial fidelity decay in the chaotic region we have found that
the strength of the perturbation used allows putting the dynamics in a perturbative region in which a Gaussian type of exponential decay is present. From Fig.5a--d we can see that the rate of the fidelity decrease depends on the perturbation strength $\Delta_\epsilon$ and has a character that differs from that found for regular dynamics.
In Fig.6a we have plotted the logarithm of the fidelity versus time square -- we can see that on this scale we can approximate the dependence via a linear function and hence, in consequence, the decay of the fidelity can be described via a function $\exp[-\mbox{const}\cdot t^2]$.

For external pumping strength high enough to put a classical system into the chaotic dynamics region (for instance $\epsilon=0.7$) increasing the perturbation strength we enter a region in which a decay of fidelity is slower than for the regular system's dynamics (see Fig.5c and 5d), and depends on the value of perturbation $\Delta_\epsilon$. We can see a characteristic decay of the fidelity (for increasing strength of perturbations) on a semilogharitmic scale. In Figs.6b and 6c we can see that the time dependence of the logarithm of the fidelity has a linear character 
and therefore, the fidelity decay can be described via a function $\exp[-A(\Delta_\epsilon)\cdot t]$ with a slope (described via $A(\Delta_\epsilon)$ - some function of $\Delta_\epsilon$) which is perturbation strength dependent.
Such a behaviour of the fidelity decay for chaotic dynamics regions is justified under the RMT theory \cite{JSB01,GPS04}.

In this paper we shall concentrate on a perturbative region, in which the rate of fidelity decay is of Gaussian type.  
Therefore, we suggest to take another parameter, which would allow us to determine whether the system's dynamics is chaotic or not, not from the analysis of a short-time fidelity decay solely but from its behaviour in long times (if compared with the Ehrenfest time for the system).

Thus, in further considerations we apply another parameter, known from the classical chaos theory -- the maximal Lyapunov exponent $\lambda_{max}$. We use it to identify regions of regular and chaotic dynamics and to confirm our conclusions arising from the fidelity evolution analysis, namely that the regions which are classically chaotic, can also be treated as chaotic ones in quantum system dynamics.
To determine the maximal Lyapunov exponent from time series we have used the procedures given in \cite{HKS99} . 
From the sign of the maximal Lyapunov exponent we have identified the region of deep classical chaos as that in which a quantum system is also chaotic (even in the classical meaning -- a positive maximal Lyapunov exponent).

For the region in which the fidelity exhibits regular oscillations, $\lambda_{max}$ estimated from the long-time series tends to zero (from negative values), whereas for
the region in which quantum beats appear for the long-time series, the maximal Lyapunov exponent tends to zero (from positive values) indicating
a quasi-periodic dynamics of the system. One should keep in mind that the region of quasi-periodic dynamics in this quantum system
can be identified in the long-time limit only. The short-time analysis totally neglects this feature.

Finally, for the regions identified earlier by the fidelity decay as being quantum-chaotic, the maximal Lyapunov exponent becomes positive. For its estimation we have used the part of the results calculated for the fidelity - we take time series which arises just after its initial decay. In this way we do not examine the fidelity decay itself but the character of its changes in the long-time regime.
For $\epsilon=0.41$ the value of the maximal Lyapunov exponent estimated from the long-time series is $\lambda_{max}\approx 0.0005$, for $\epsilon=0.505$ it becomes $\approx 0.0047$ but for $\epsilon=0.8$ its value
is $\approx 0.009$. This means that indeed, a quantum system for these values of kick strength is chaotic in a classical sense -- we have obtained a positive maximal Lyapunov exponent from the time series corresponding to the quantum parameter evolution describing the dynamics of the system. 

As follows, the long-time behaviour of the fidelity can be treated as an
indicator of quantum chaos. When the classical system considered is in the region of classically deep chaos, its quantum counterpart is chaotic too.

\subsection{\bf Fidelity-like parameter ${ F}_N$}
It is known that when the Kerr-like oscillator is externally "kicked"
by a series of ultra-short pulses and the strength of these "kicks" can induce the
regular dynamics of the oscillator, $n$-photon Fock
states are generated. It has been already shown \cite{L96,LT94} that under some assumptions the system can be
treated as a one-photon state generator. 
But, when the system behaves chaotically, the states with greater and greater number of photons are involved -- see Fig.7c. At this point
we propose another fidelity-like measure ${F}_N$  that would explicitly include
the energy of the system. It should be noted that the classical bifurcation diagram gives us the information about the classical energy of the system. Hence, the fidelity-like parameter ${F}_N$ we propose here, comprises the operator of the mean number of photons and can be expressed as:
\begin{equation}
{ F}_N(t)=|\langle\Psi(t=0)|\left(U^{+}_{Kp}U^{+}_{NL}\right)^k \hat{a}^{+}\hat{a} \left(U_{NL}U_K\right)^k|\Psi(t=0)\rangle|\,\,.
\label{eq9}
\end{equation}

We can expect that the new fidelity-like parameter ${F}_N$, including a drastic
increase in the mean system's energy, would be especially useful for investigation of the regions of chaotic dynamics. However, for the cases of regular dynamics
this parameter should exhibit the regular character of evolution. 
Indeed, from Fig.7a we can easily see such a behaviour.

The fidelity-time dependence is now a composition of two frequencies - we can easily see the effect of their superposition and some beats occur in the evolution of $F_N$. 
The quick regular revivals of the parameter $F_N$ are influenced via slower oscillations with the frequency equal to that of the fidelity $F$ revivals --- see Fig.7a.

Moreover, while the number of photons
increases significantly, the time-dependence of
our fidelity-based parameter $F_N$ is also influenced by these
changes -- see Fig.7b. When we take a closer look at $F_N$ on the short time scale (the inset in Fig.7b), we can observe rapid oscillations of small amplitude (resulting from the quick changes in the number of photons). These oscillations are preserved during the whole time of the system's evolution.

It should be emphasised that for
the regions corresponding to both the quasi-periodic dynamics and
the classical "deep"
chaos the evolution of the quantum parameters
${F}$ as well as  ${F}_N$ exhibits chaotic behaviour. 

\section{The entropic measure ${\cal E}$.}
To measure the changes in fidelity between quantum
states we shall propose an entropic parameter ${\cal E}$ based on the
fidelity evolution and describing its
character. Having defined $F(t)$ one could find the Fourier transform of this function as follows:
\begin{equation}
{\cal F}(\omega)=\sum\limits_{t=t_{min}}^{t_{max}}F(t)e^{-i\omega t} \,\,\,,
\label{eq10}
\end{equation}
and hence, the power spectrum  ${\cal P}(\omega)=|{\cal
  F}(\omega)|^2$. Then we normalise it to get ${\cal
  P}_N(\omega)$ and finally, we define the "entropy" of changes in fidelity as follows:
\begin{equation}
{\cal E}=-\sum\limits_{\omega}{\cal P}_N(\omega)\log\left({\cal P}_N(\omega)\right) \,\,\, .
\label{eq11}
\end{equation}  
Obviously, one should keep in mind that due to the discrete character of the kicked system evolution these parameters have been defined as sums, not integrals.

Thus, Fig.8 shows changes in the newly defined measure with the external field strength $\epsilon$.
As bifurcation diagrams allow identifying the regions of regular and chaotic dynamics for a classical system, 
the character of changes in the fidelity is described by the entropic parameter (\ref{eq11}).

When the oscillator's dynamics is regular, ${\cal E}$ smoothly changes its value with $\epsilon$. This corresponds to the region of the maximal Lyapunov exponent tending to zero from negative values and therefore, it can be identified as that of the regular system dynamics.
For long times  we have already seen that there is also a region where some quantum beats can be observed (for this case the maximal Lyapunov exponent tends to zero from positive values). We can see from Fig.6 that this situation corresponds to the region of a rapid growth in the entropy. Thanks to this fact ${\cal E}$ can be used as an indicator of the quantum chaos border.
Finally, irregular changes in the entropy (after some growth of its value) appear in the region of the positive maximal Lyapunov exponent. We recognize this region as that of quantum chaos. Moreover, we see that all the characteristic features of the quantum system dynamics considered are clearly manifested on the entropy  dependence on $\epsilon$ and therefore,  $\cal E$ can also be used as indication of the quantum chaotic behaviour of the system discussed.

Additionally, we can say that the chaotic region surrounded by regular ones (for $\epsilon\approx 0.35$) is
characteristic only of the classical dynamics of the
oscillator. There are no changes in the quantum indicators of chaos for these $\epsilon$ values.

\section{Conclusions}
We have presented the application of a long-time analysis of the fidelity between two quantum states
for determination whether the quantum system exhibits quantum-chaotic behaviour or not.
For our considerations we have chosen the system of a
Kerr-like  oscillator externally driven by a series of ultra short
coherent  pulses \cite{L96,LT94} -- such a system can exhibit regular or chaotic behaviour. 
To investigate the system's dynamics  we
have used a strictly quantum parameter, namely, the fidelity between the
quantum  state generated $|\Psi_u(n)\rangle$ and the state slightly
perturbed $|\Psi_p(n)\rangle$. We have followed the ideas presented in
\cite{P84,WLT02,EWLC02}, where the time-dependence of the fidelity was used
to analyse the system's dynamics. 
As has been shown in these papers, for the short-time case,
the fidelity $F$ decreases exponentially in the chaotic region. We have analysed this decay in order to find whether it can be comparable with the results obtained from the RMT theory. We have came to the conclusion that as the perturbation strength is such that the Gaussian type of the fidelity decay is present, the chaocity (or not) of the system's dynamics can be concluded from another parameter.
Contrary to the discussion presented in \cite{WLT02,EWLC02}, in this paper we have analysed the long-time  fidelity time-evolution. Moreover, we have proposed not only the fidelity-like parameter $F_{N}$ comprising the information concerning the mean number of photons in the system, but also some entropic measure $\cal E$ that, as we have shown, can be an indicator of quantum chaos too.

We have shown that when the system is close to the quantum chaotic border, the long-time analysis gives the clearly seen modulations of the previously periodically oscillating fidelity. Moreover, for this case the maximal Lyapunov exponent (known from the classical chaos analysis) tends to zero from positive values, indicating a quasi-periodic dynamics.
In the region corresponding to the "deep" classical chaos, the fidelity exhibits significant irregular changes (the short-time fidelity decay ensures us that this is a quantum chaotic region as well).
This means that for this case we cannot specify the quantum state which is generated
in the  process of interaction with external field. This feature is characteristic
of the evolution of the chaotic nature. Moreover, we have shown that the irregular changes are clearly visible for the evolution of newly defined fidelity-like  $F_N$ parameter (that is of quantum nature as well). In addition, the entropic parameter $\cal E$ based on the fidelity $F$ proposed here, changes its value rapidly as the coupling strength $\epsilon$ corresponds to the region of classically chaotic dynamics. The maximal Lyapunov exponent in this region is positive indicating chaotic behaviour of the system as well.

As problems dealing with quantum chaotic systems are still explored,
we believe  that the considerations of the long-time behaviour of
fidelity and the analysis of the fidelity-like  parameter $F_N$  can be an interesting point for further investigation in
the  field of quantum chaos.
Moreover, we believe that the entropic parameter $\cal E$ can be a useful tool for indication of the quantum-chaotic behaviour as well.

\newpage
 \begin{figure}
 \begin{center}
\hspace*{-1cm}\resizebox{16cm}{10cm}
                {\includegraphics{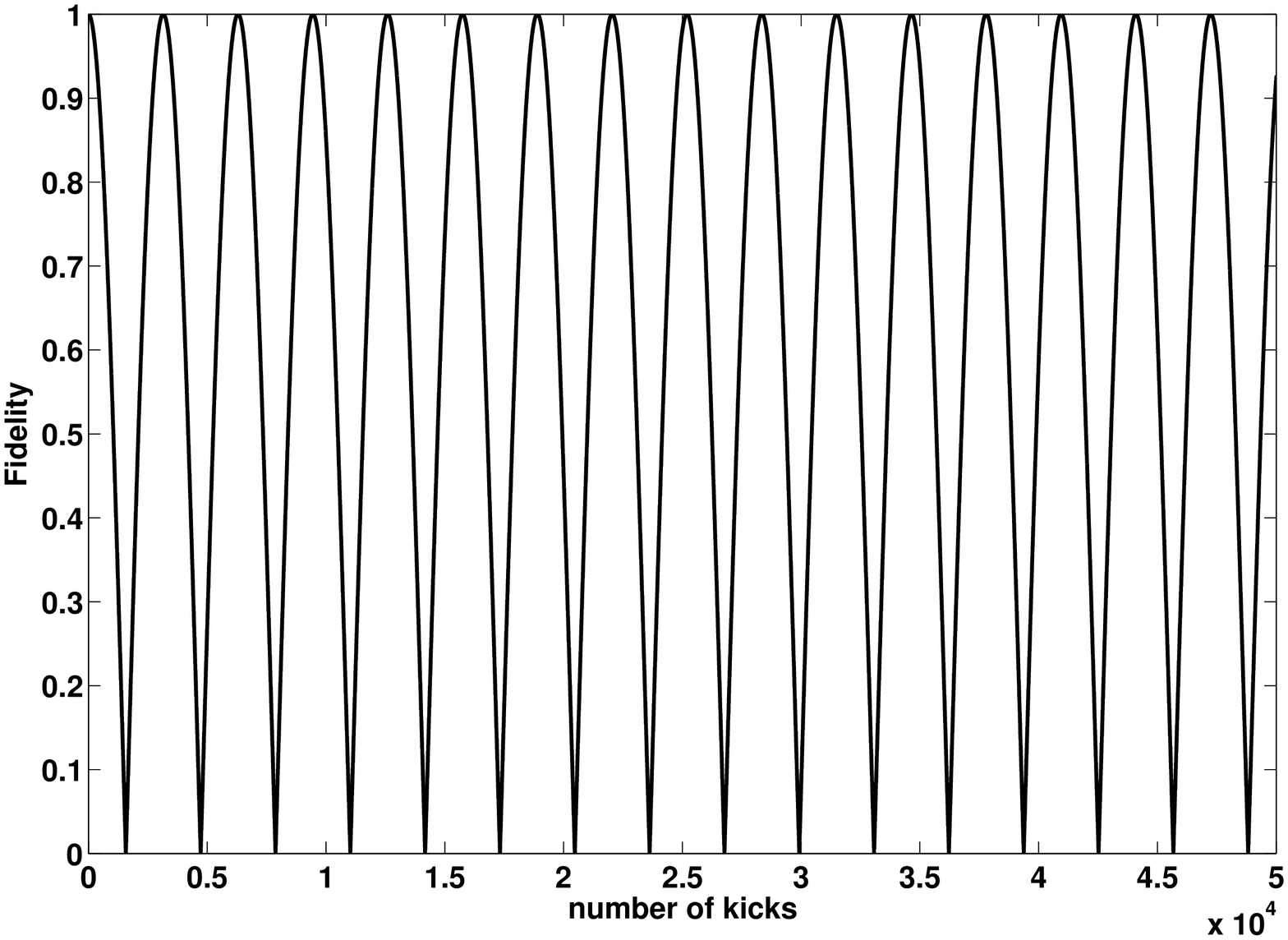}}
\end{center}
\caption{The fidelity $F$ versus the number of kicks for $\epsilon=0.1$ and $\Delta_\epsilon=0.001$. 
All the energies are measured in units of $\chi$. The time 
$T=\pi$, the nonlinearity parameter $\chi=1$.}
\end{figure}
\newpage
 \begin{figure}
 \begin{center}
\hspace*{-1cm}\resizebox{11.5cm}{7.cm}
                {\includegraphics{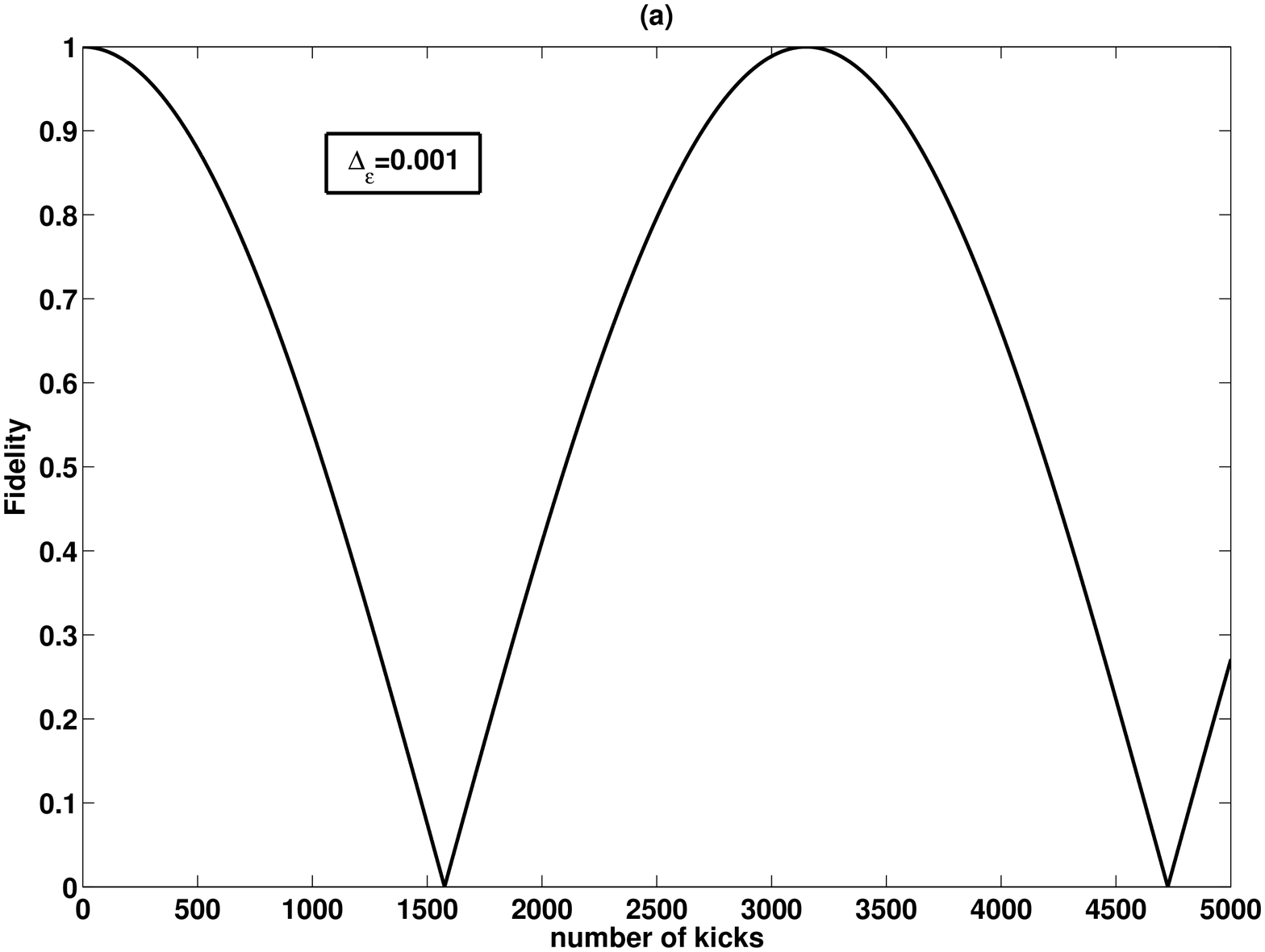}}
\hspace*{-1cm}\resizebox{11.5cm}{7.cm}                
                {\includegraphics{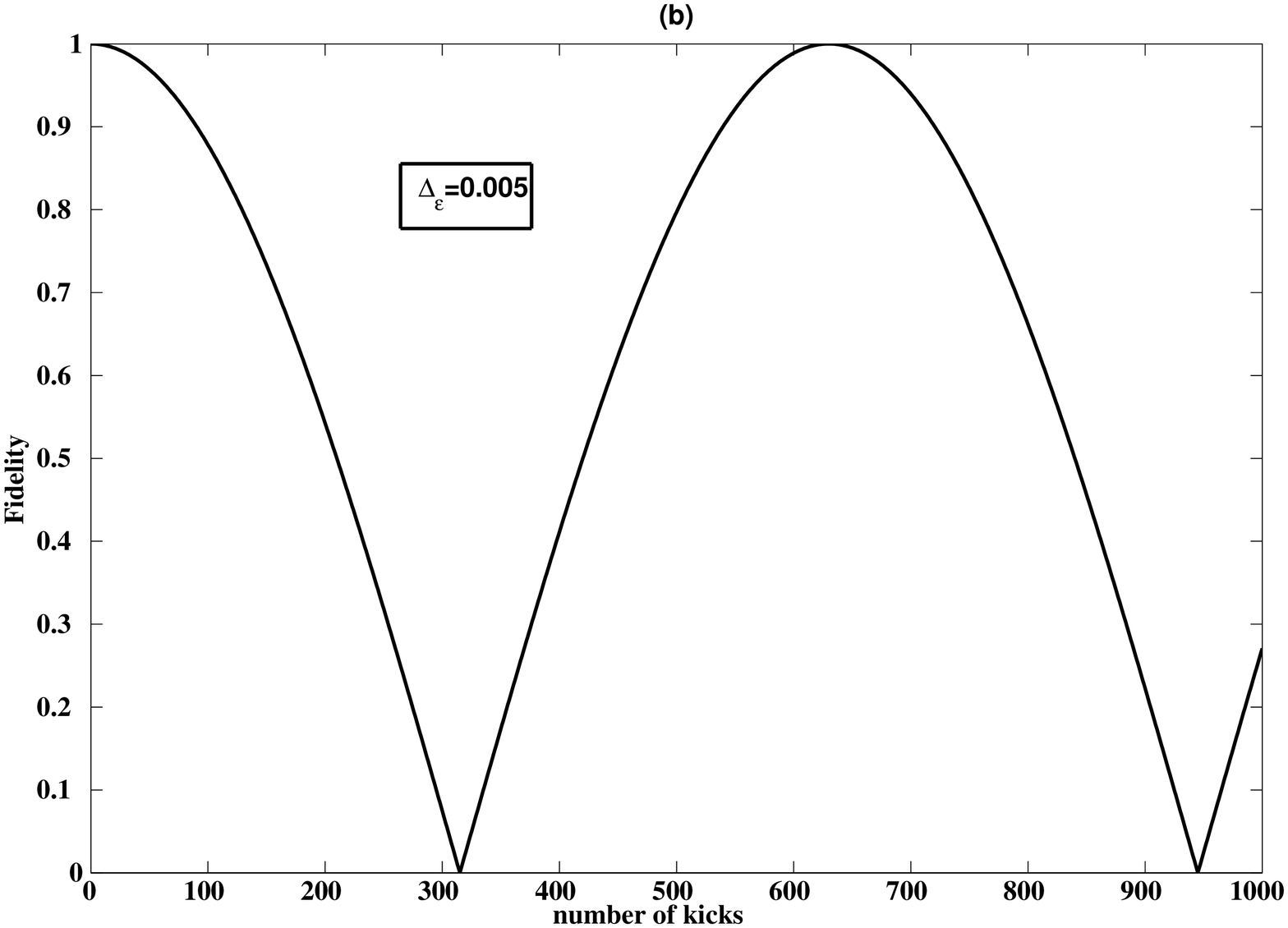}}
\hspace*{-1cm}\resizebox{11.5cm}{7.cm}                
                {\includegraphics{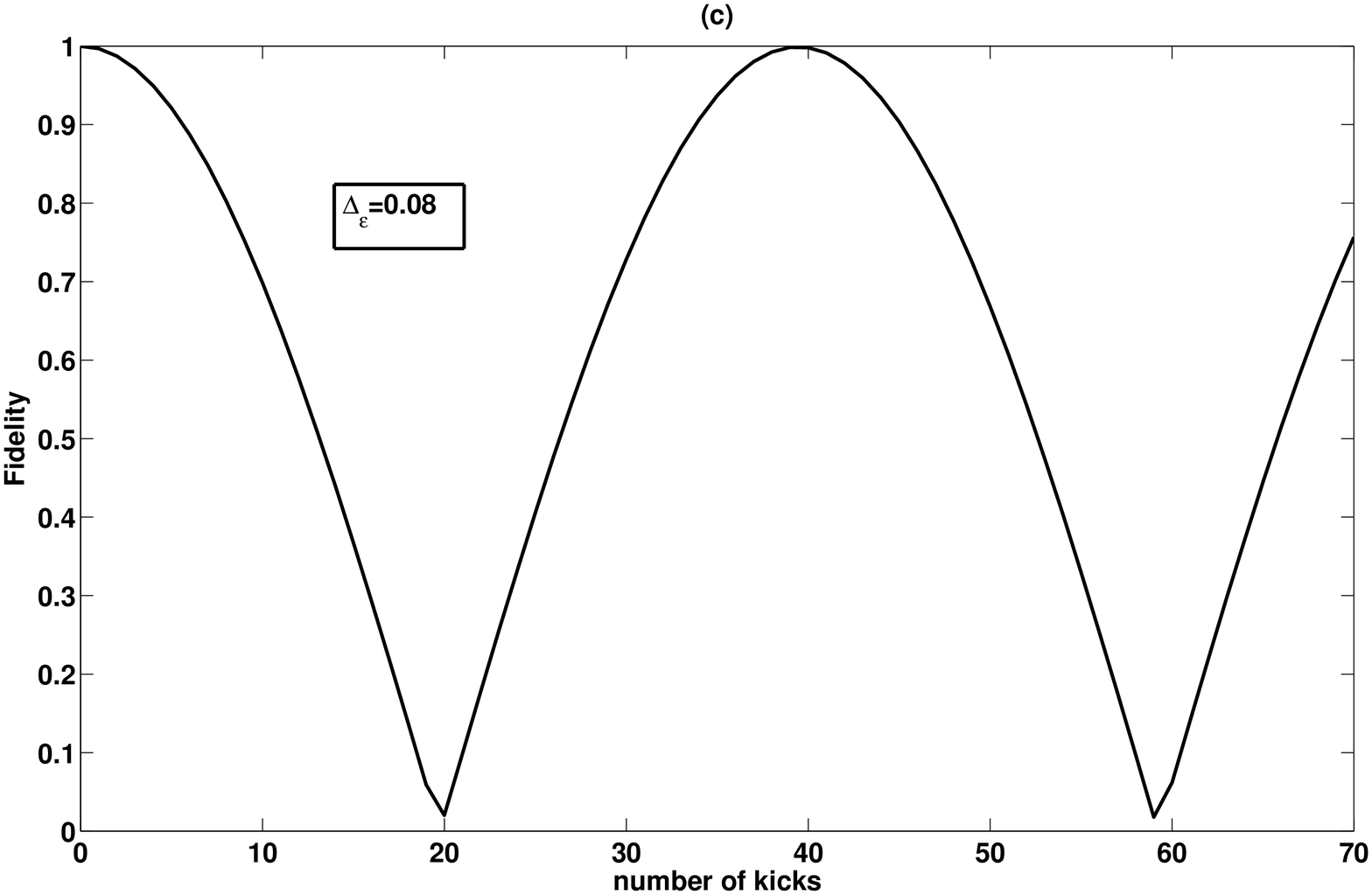}}
\end{center}
\caption{The fidelity $F$ versus the number of kicks for $\epsilon=0.1$ and $\Delta_\epsilon=0.001$ 
--- (a). At (b) for $\Delta_\epsilon=0.05$  and  $\Delta_\epsilon=0.08$ at (c). The remaining parameters are the same as in Fig.1.}
\end{figure}
\newpage
 \begin{figure}
 \begin{center}
\hspace*{-1cm}\resizebox{16cm}{10cm}
                {\includegraphics{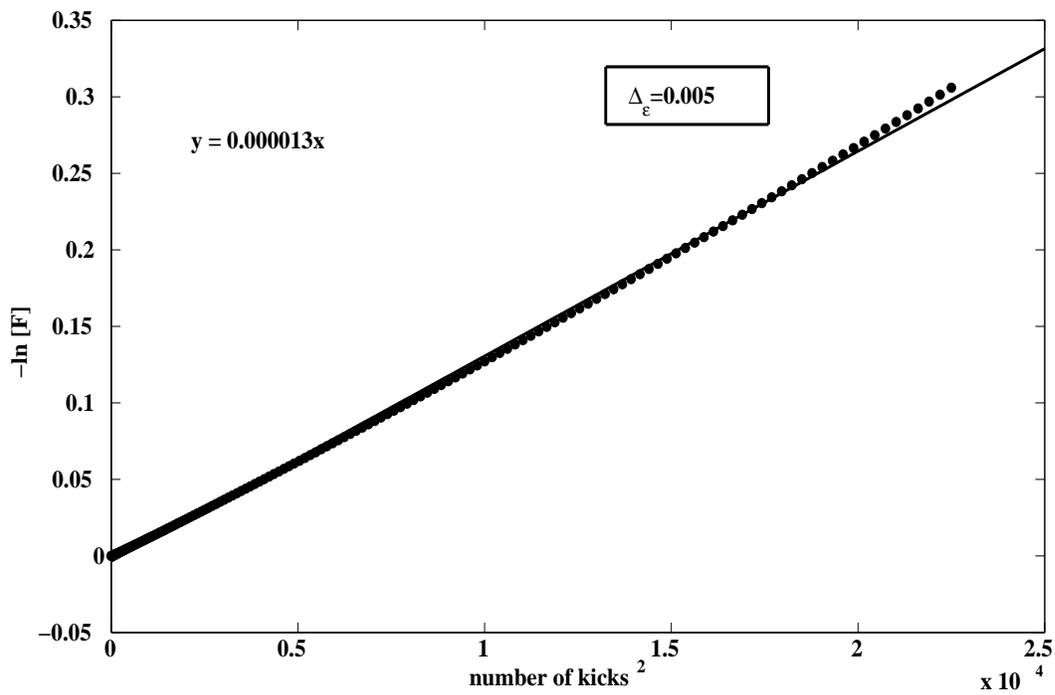}}
\end{center}
\caption{The logarithm of fidelity $F$ versus the squared number of external kicks for $\epsilon=0.1$ and $\Delta_\epsilon=0.05$. The remaining parameters are the same as in Fig.1.}
\end{figure}
\newpage
 \begin{figure}
 \begin{center}
\hspace*{-1cm}\resizebox{12cm}{8cm}
                {\includegraphics{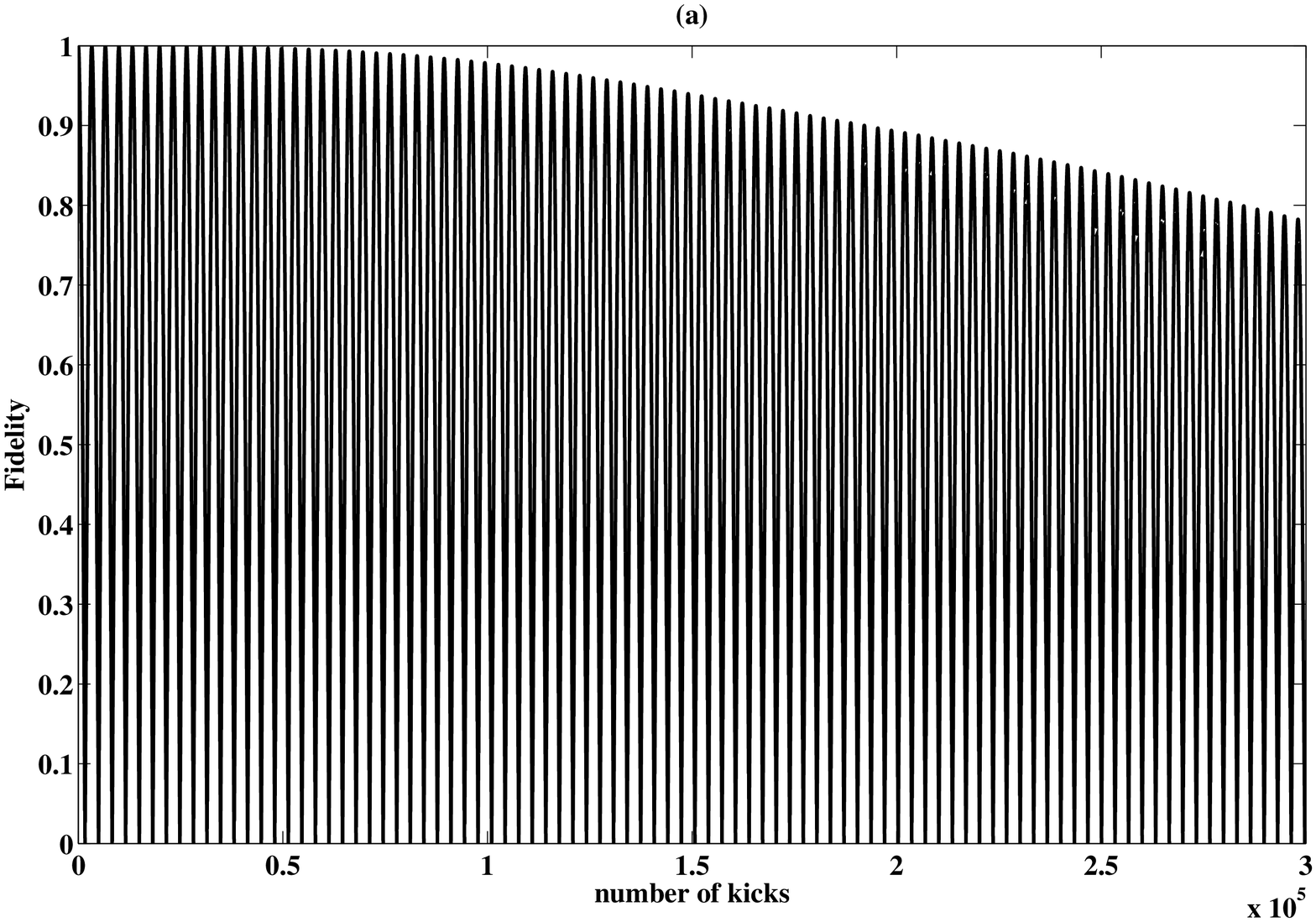}}
\hspace*{-1cm}\resizebox{12cm}{8cm}                
                {\includegraphics{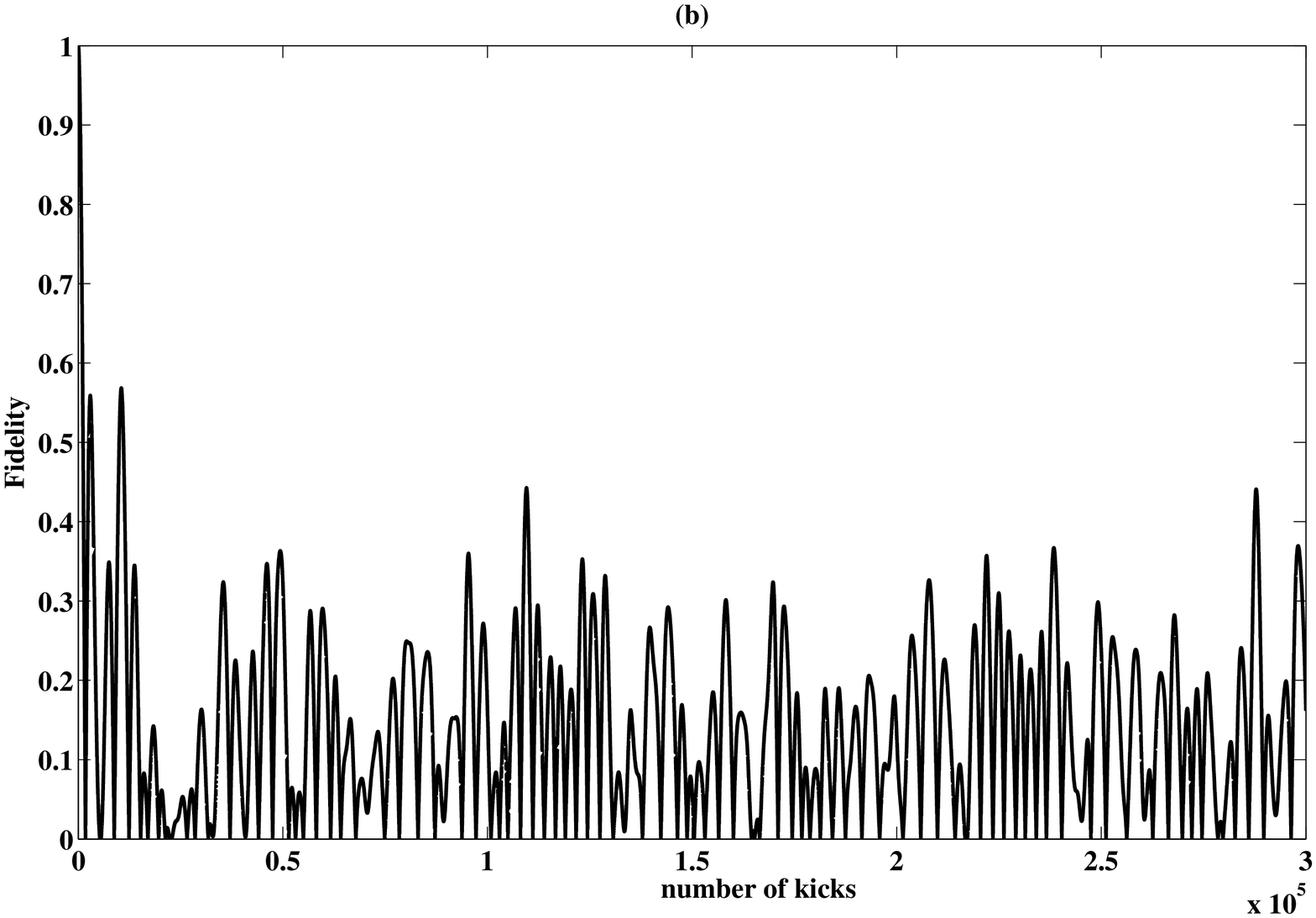}}

\end{center}
\caption{The fidelity $F$ versus the number of kicks for various values of coupling strength: (a)    $\epsilon=0.385$,  (b)  $\epsilon=0.505$. The remaining parameters are the same as in Fig.1.}
\end{figure}
\newpage
 \begin{figure}
 \begin{center}
\hspace*{-1cm}\resizebox{12cm}{8cm}
                {\includegraphics{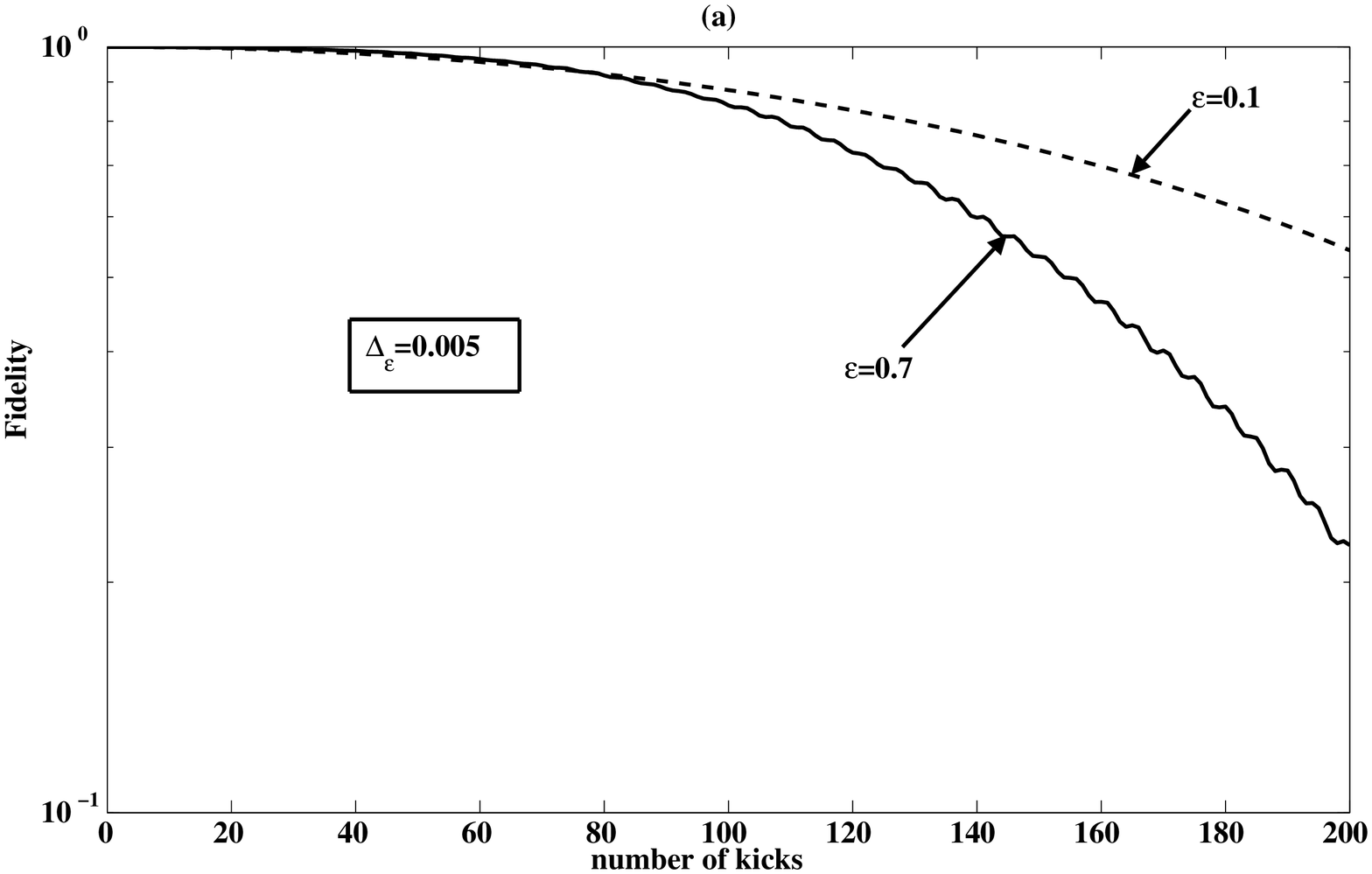}\includegraphics{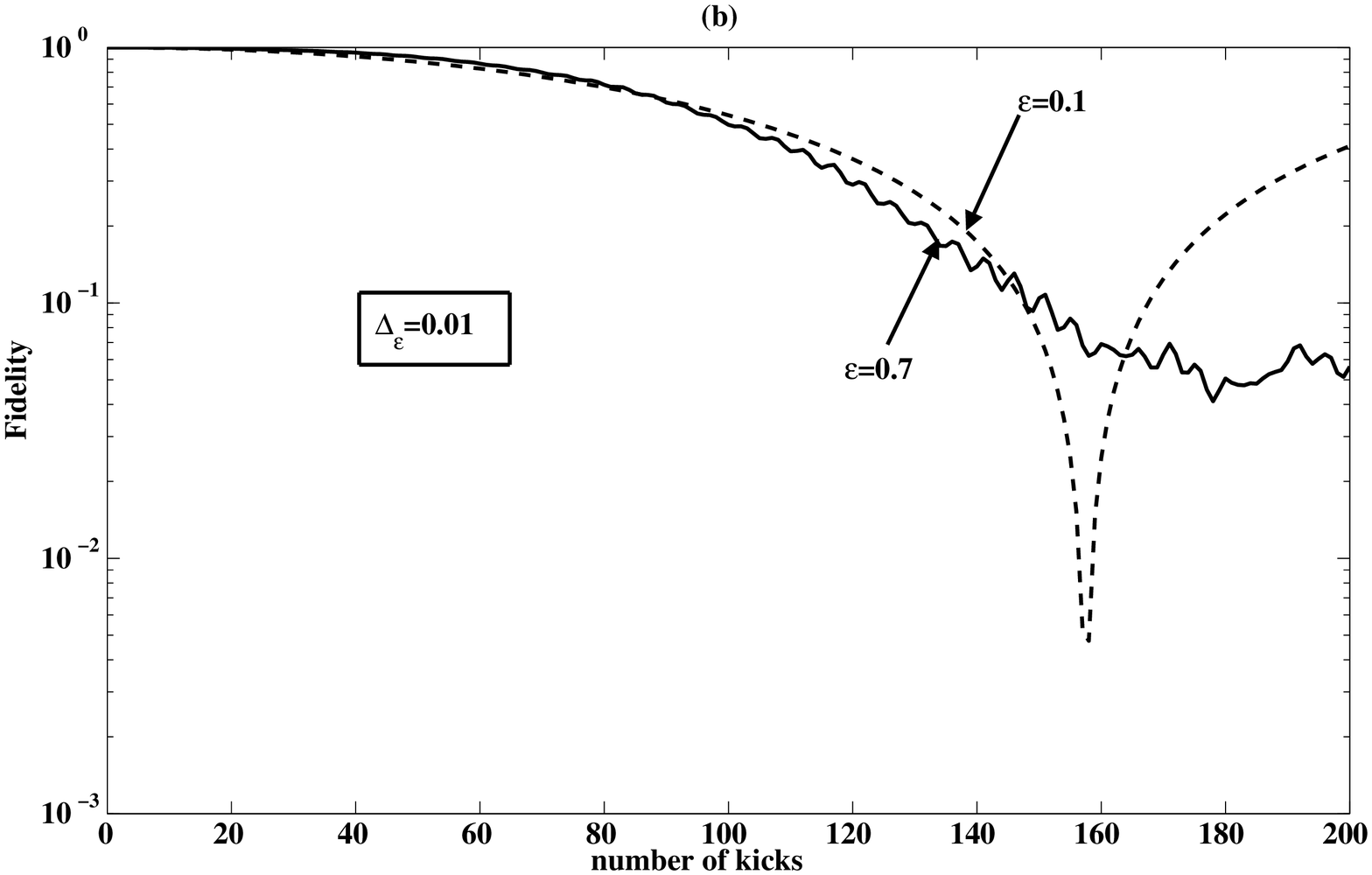}}
\hspace*{-1cm}\resizebox{12cm}{8cm}                
                {\includegraphics{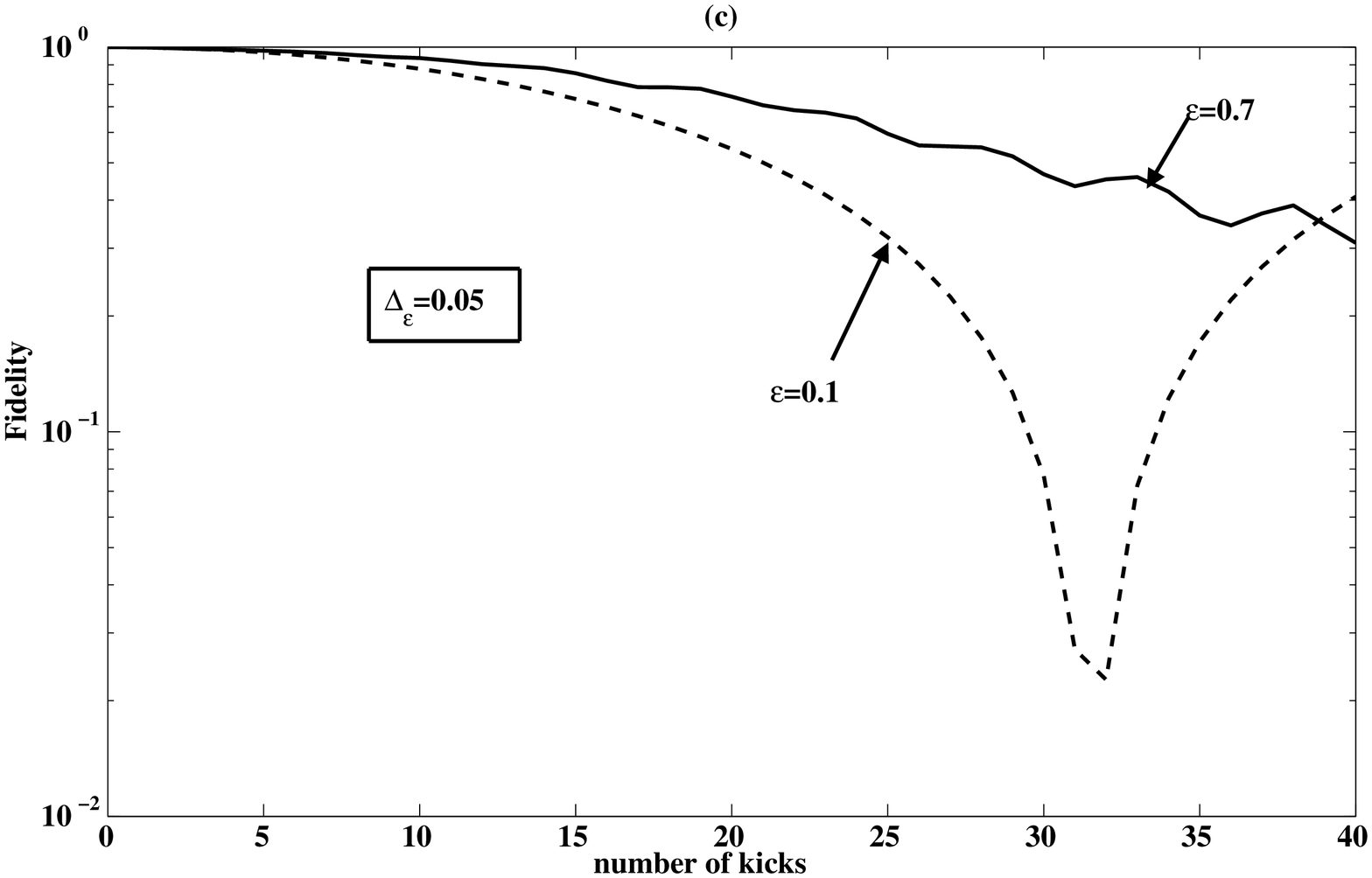}\includegraphics{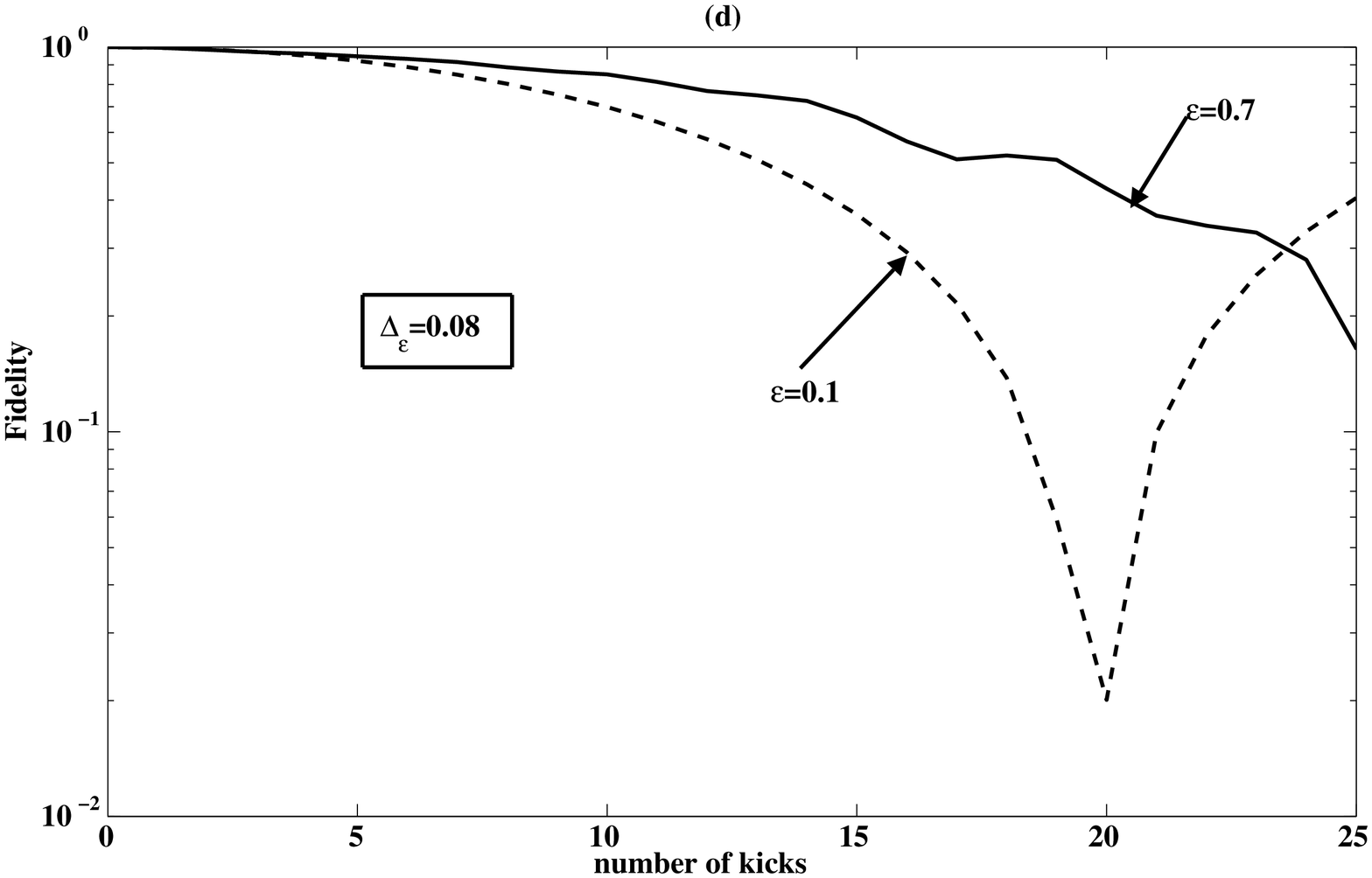}}

\end{center}
\caption{The fidelity $F$ versus the number of kicks for various values of perturbation strength at a semi-logarithmic scale: (a) $\Delta_{\epsilon}=0.005$; (b) $\Delta_{\epsilon}=0.01$; (c) $\Delta_{\epsilon}=0.05$ and (d) $\Delta_{\epsilon}=0.08$. The remaining parameters are the same as in Fig.1.}
\end{figure}
\newpage
 \begin{figure}
 \begin{center}
\hspace*{-1cm}\resizebox{12cm}{7cm}
                {\includegraphics{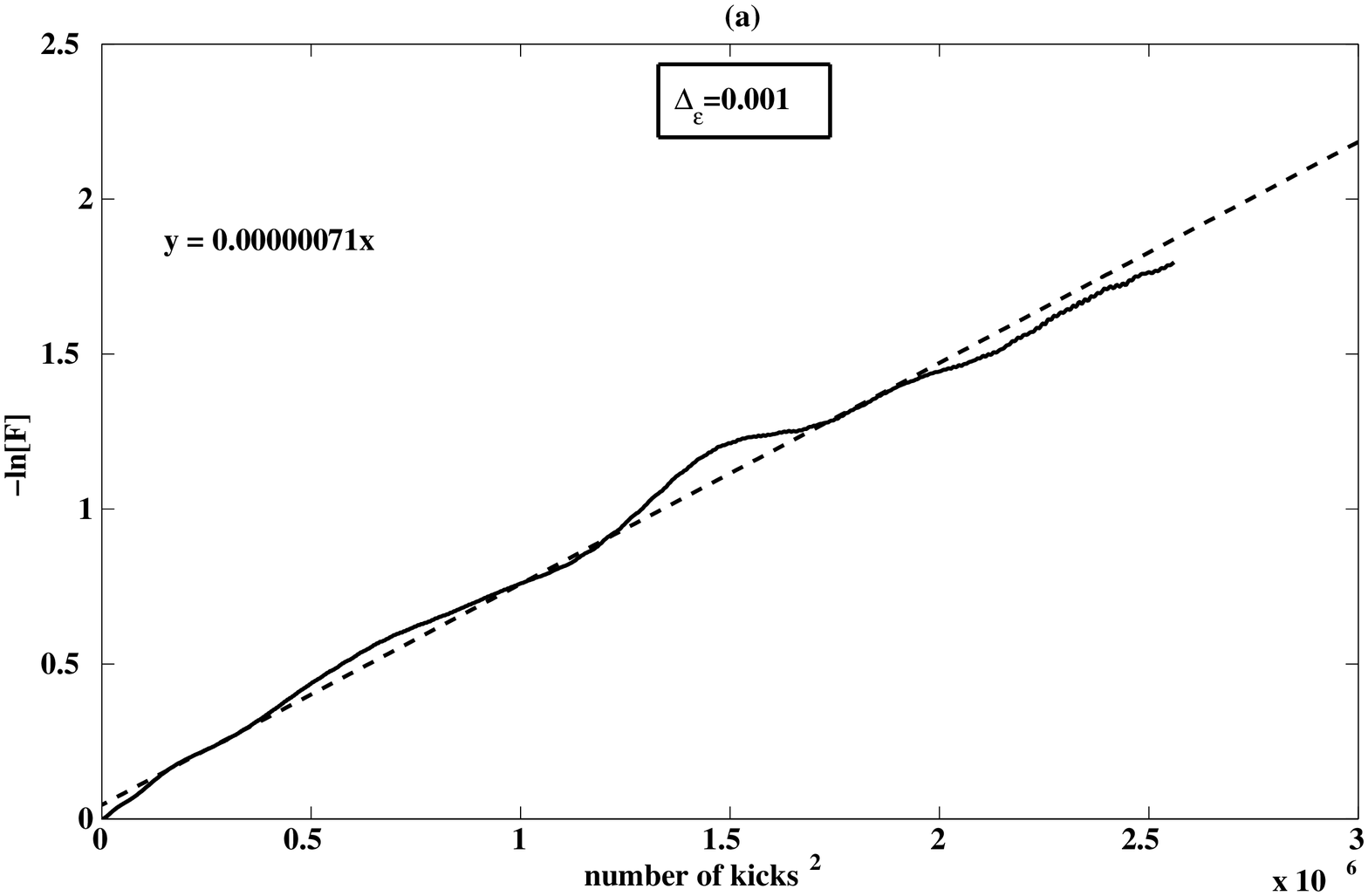}}
\hspace*{-1cm}\resizebox{12cm}{7cm}                
                {\includegraphics{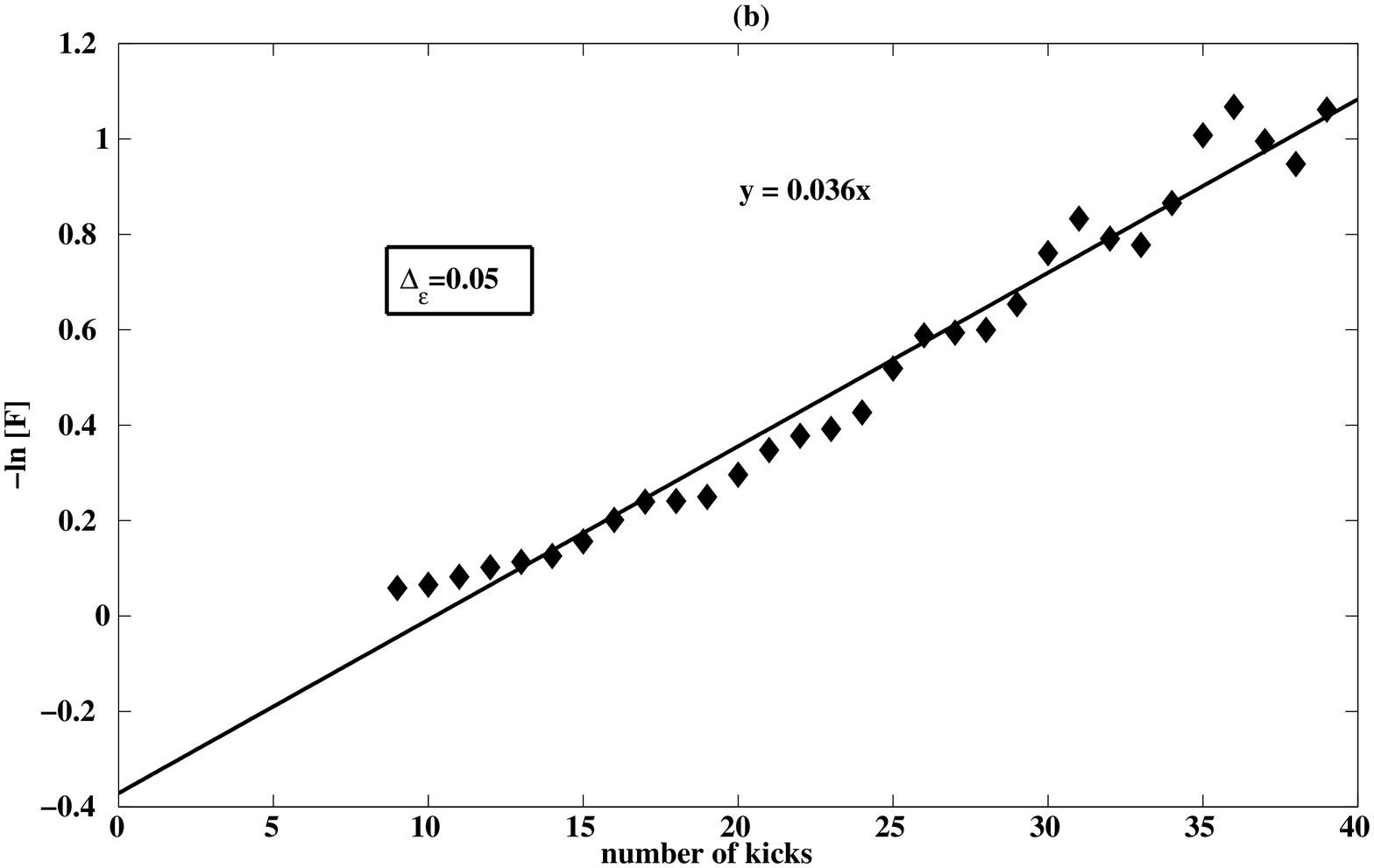}}
\hspace*{-1cm}\resizebox{12cm}{7cm}                
                {\includegraphics{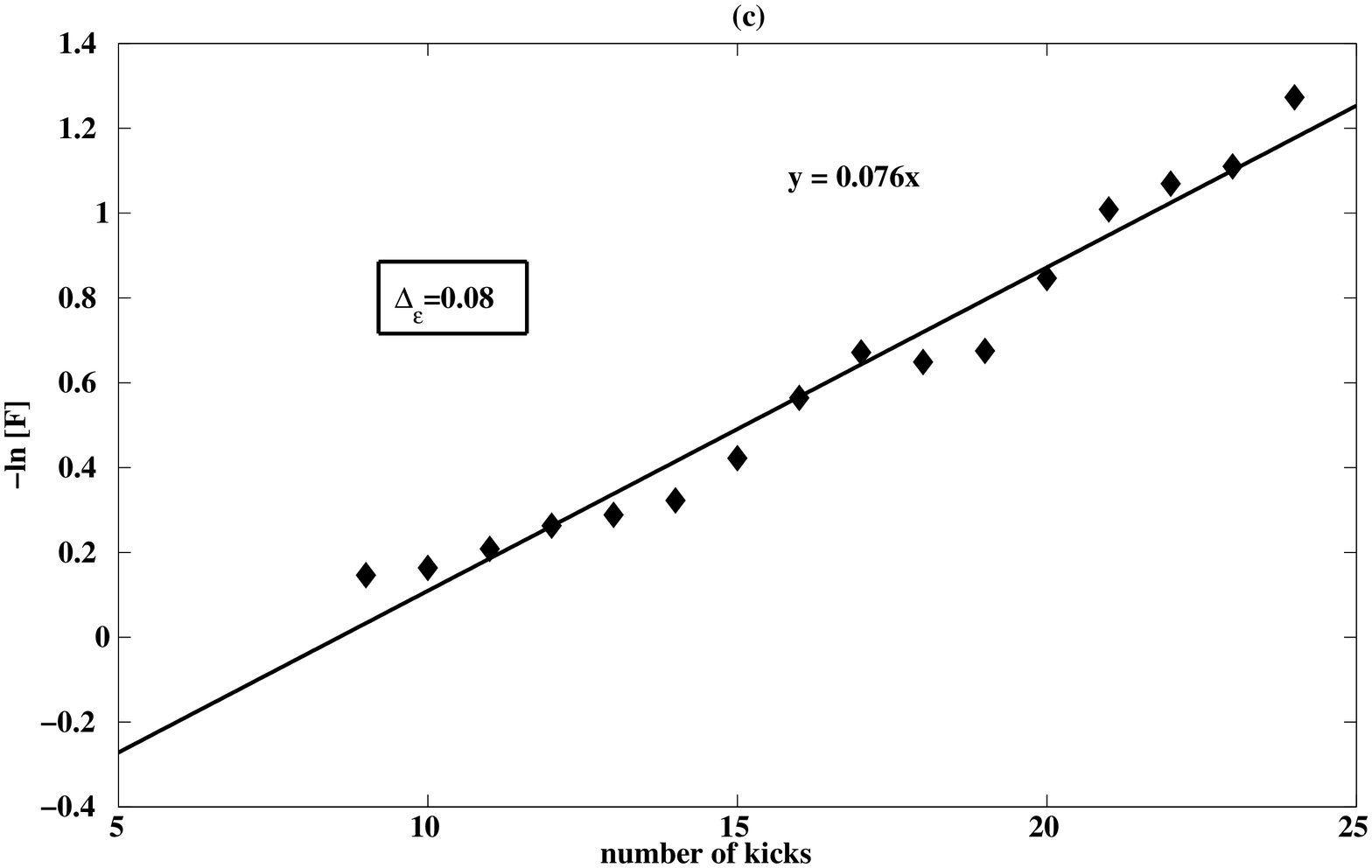}}
\end{center}
\caption{$Log[F(t)]$ for $\epsilon=0.7$ (a) versus the number of kicks square for $\Delta_{\epsilon}=0.001$;
 versus the number of kicks but for $\Delta_{\epsilon}=0.05$ at (b) and $\Delta_{\epsilon}=0.08$ at (c).
The remaining parameters are the same as in Fig.1.}
\end{figure}
\newpage
 \begin{figure}
 \begin{center}
\hspace*{-1cm}\resizebox{12cm}{7cm}
                {\includegraphics{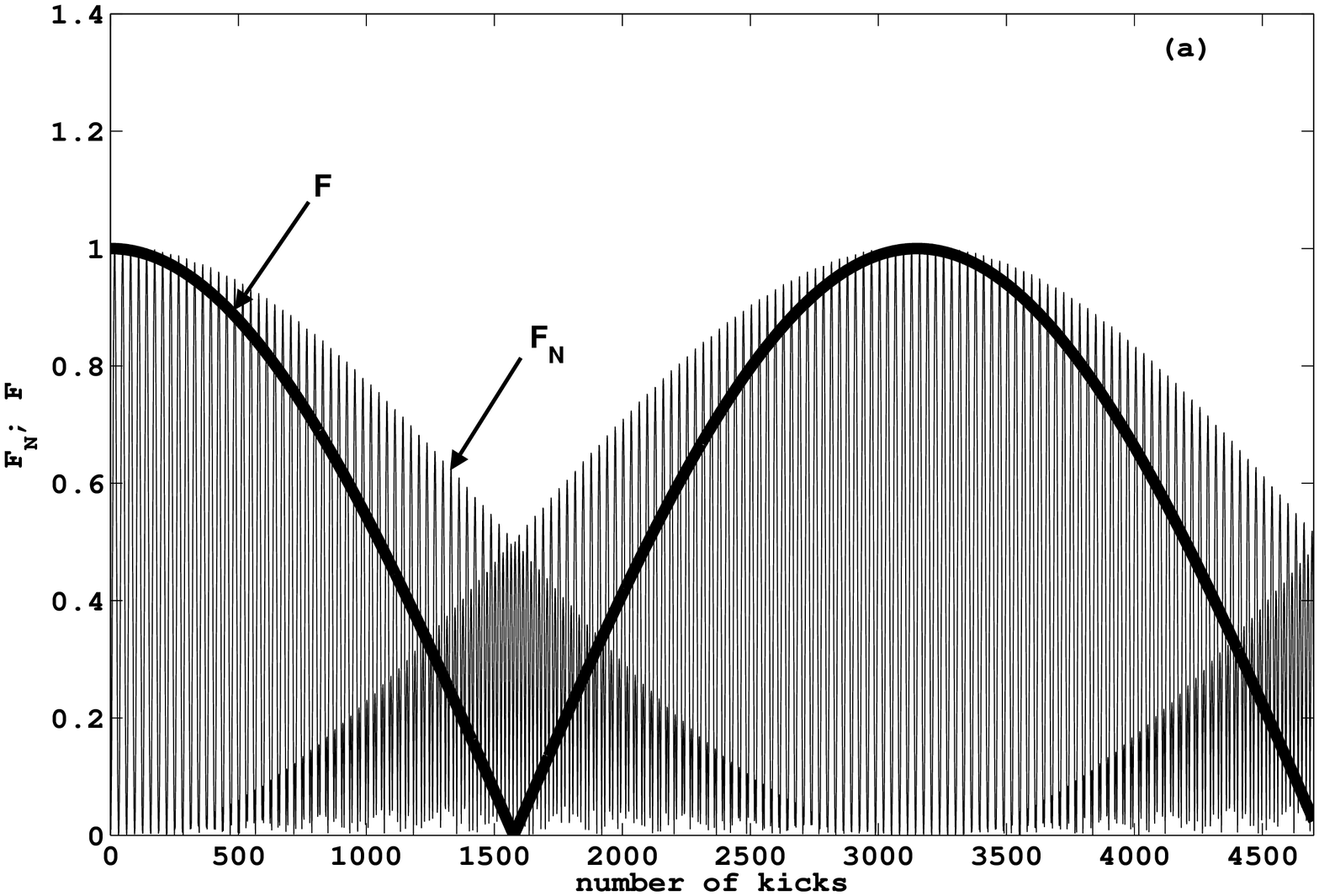}}
\hspace*{-1cm}\resizebox{12cm}{7cm}                
                {\includegraphics{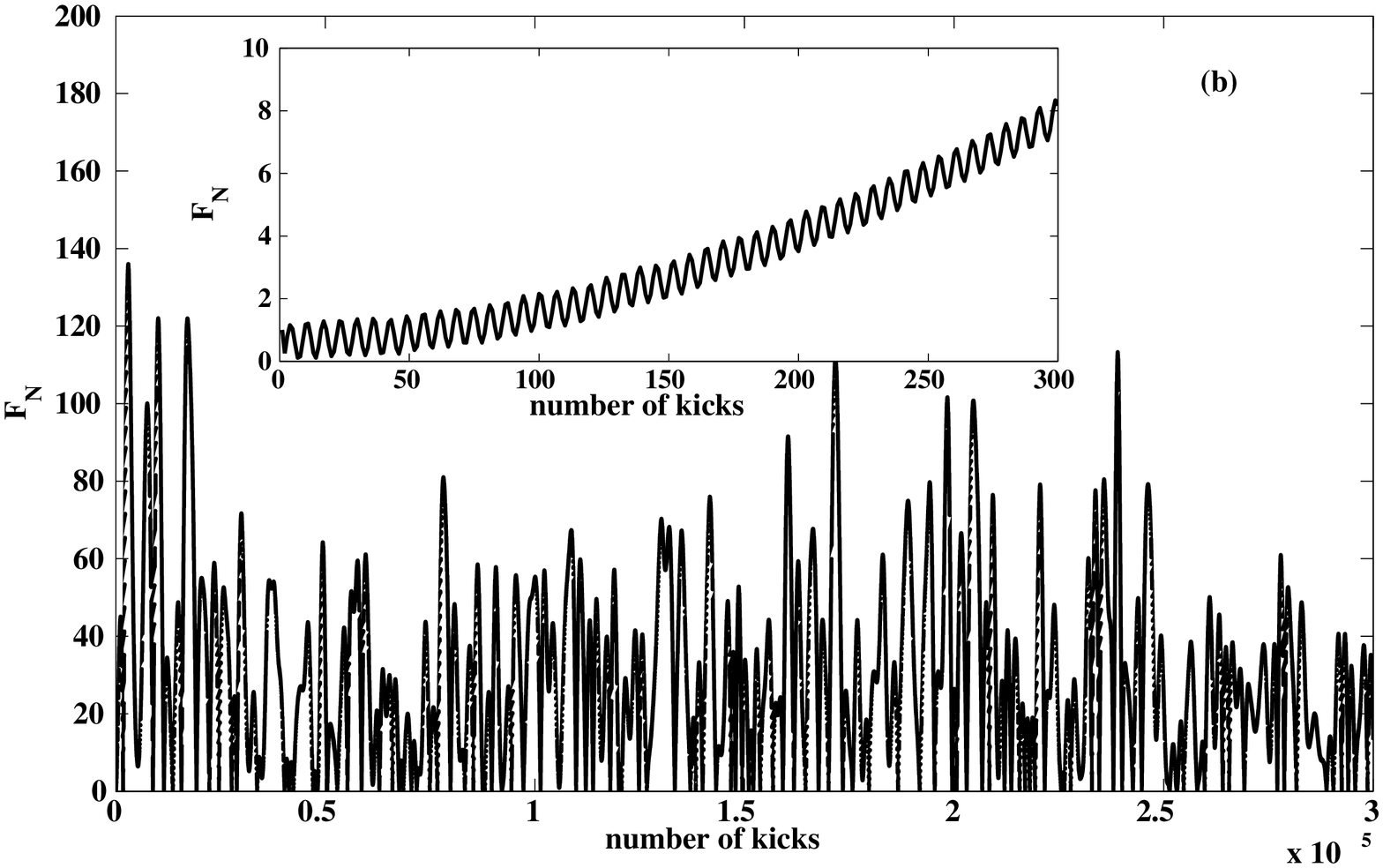}}
\hspace*{-1cm}\resizebox{12cm}{7cm}                
                {\includegraphics{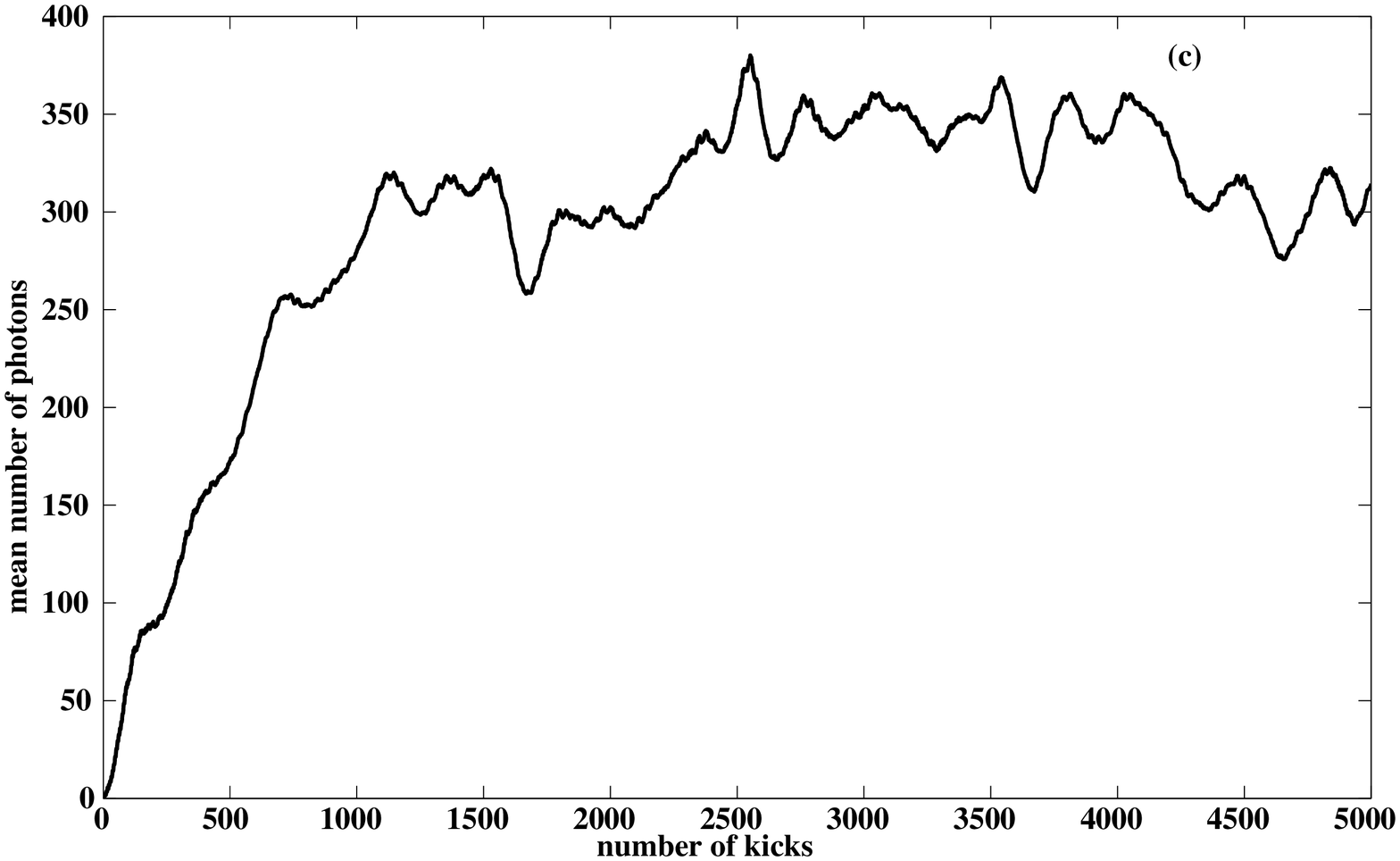}}
\end{center}
\caption{The fidelity $F_N$ versus the number of kicks for: (a) $\epsilon=0.1$ (for comparison, thick solid line is $F(t)$); (b) $\epsilon=0.505$ (in the inset we have shown $F_N$ on a different time scale). Fig.c -- the mean number of photons $\langle a^{+}a\rangle$ for $\epsilon=0.505$.
All remaining parameters are indentical to those from Fig.1.}
\end{figure}
\newpage
 \begin{figure}
 \begin{center}
\hspace*{-1cm}\resizebox{16cm}{10cm}
                {\includegraphics{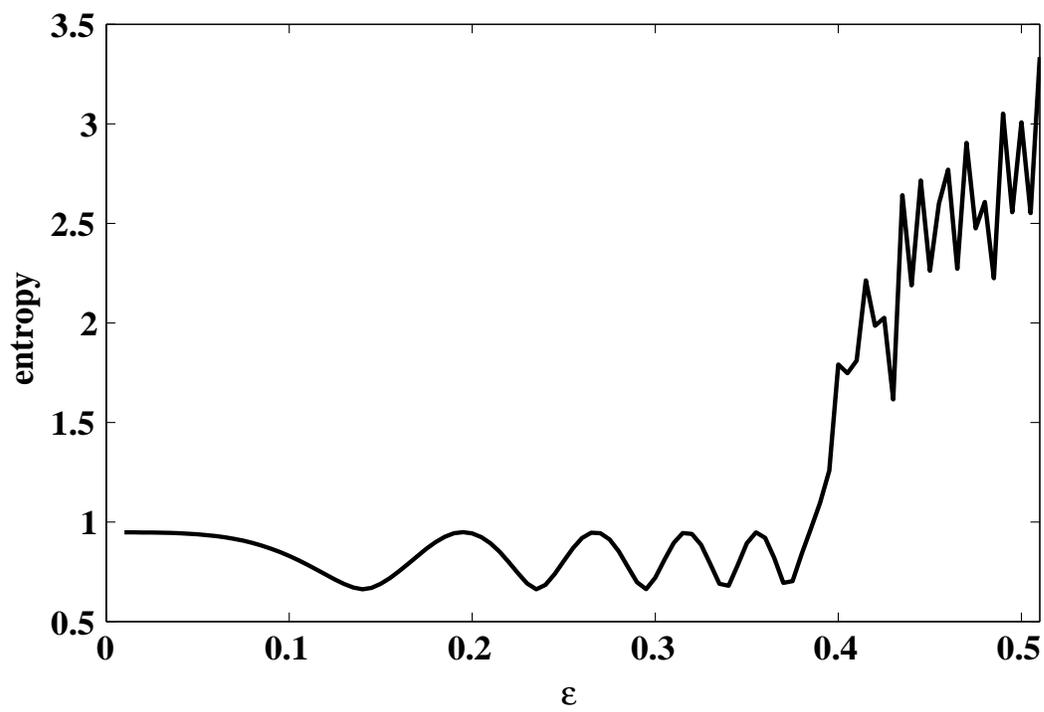}}
\end{center}
\caption{The entropy ${\cal E}$  vs. the interaction strength $\epsilon$. 
The parameters are the same as in Fig.1.}
\end{figure}

\end{document}